\documentclass[fleqn]{article}
\usepackage{graphicx}

\begin{document}

\title{
{\sf   Estimate of the Three-Loop Perturbative Contribution 
              to Inclusive Semileptonic $b\rightarrow   u$ Decays   } }

\author{
M.R. Ahmady\thanks{email: mahmady2@julian.uwo.ca}, F.A. Chishtie\thanks{email: fachisht@julian.uwo.ca}, 
V.Elias\thanks{email: velias@julian.uwo.ca}\\
{\sl Department of Applied Mathematics}\\
{\sl University of Western Ontario}\\
{\sl London, Ontario N6A 5B7, Canada.}\\[10pt]
 T.G. Steele
\thanks{email: steelet@sask.usask.ca}
\\
{\sl Department of Physics and Engineering Physics}\\
{\sl University of Saskatchewan}\\
{\sl Saskatoon, Saskatchewan S7N 5E2, Canada.}
}

\maketitle
\begin{abstract}
We utilize asymptotic Pad\'e-approximant methods to estimate the three-loop order $\overline{{\rm MS}}$-scheme coefficients within the inclusive $b\rightarrow  u \bar\nu_\ell\ell^-$ decay rate for four and five active quark flavours.  The estimates we obtain for the three renormalization-group-accessible coefficients within the three-loop contribution are all found to be within 5.1\% of their true values, using a least-squares procedure in conjunction with an asymptotic Pad\'e-approximant estimate of the three-loop term over the entire $\mu \ge m_b$ domain. Given the input values $\alpha_s(M_Z) = 0.118 \pm 0.004$ and $m_b(m_b) =4.17 \pm 0.05\, {\rm GeV}$, the three-loop expression for the purely-perturbative contribution to the $b \rightarrow u\bar\nu_\ell\ell^-$ decay rate is minimally sensitive to renormalization scale at $\mu = 1.775\, {\rm GeV}$, at which scale the three-loop contribution is estimated to be only 1.4\% of the leading tree-order contribution. We estimate the full perturbative decay rate to be $192\pi^3\Gamma(b\rightarrow u\bar\nu_\ell\ell^-)/\left(G_F^2\vert V_{ub}\vert^2\right) = 2070 \,{\rm GeV^5} \pm 16\%$, inclusive of theoretical uncertainties from series truncation, the input parameters, and the estimation procedure.
\end{abstract}
One of the outstanding issues in $B$ physics is the accurate determination of
the CKM matrix element $V_{ub}$, whose magnitude corresponds to one side of the
unitarity triangle.  Experimental determination of the angles within  this triangle is anticipated soon from $B$-factories, {\it i.e.} BELLE and BaBar detectors, and from experiments
by the CDF and CLEO III collaborations.  A more precise determination of $\vert V_{ub}\vert$ can test the
consistency of of the CKM picture of quark mixing within the standard model.  The value of $\vert V_{ub}\vert$
can be extracted from the inclusive semileptonic width $\Gamma(B\rightarrow X_u \bar\nu_\ell \ell^-)$ once a more
accurate determination of this decay rate becomes available.  In order to 
reduce the theoretical uncertainty in this rate, it is useful to determine its
purely perturbative QCD corrections as accurately as possible. 

    In perturbative QCD, the inclusive semileptonic $b\rightarrow   u$ decay rate
$\Gamma(b\rightarrow u\bar\nu_\ell \ell^-)$
 may be expressed as \cite{vanRit}
\begin{eqnarray}
\frac{1}{K}\Gamma\left(\mu, m_b(\mu),x(\mu)\right)=m_b^5(\mu)
\biggl(1&+&\left[a_0-a_1\log(w)\right] x(\mu)+
\left[ b_0-b_1\log(w)+b_2\log^2(w)\right] x^2(\mu)
\nonumber\\
& +&\left[ c_0-c_1\log(w)+c_2\log^2(w)-c_3\log^3(w)\right] x^3(\mu)
+\ldots \biggr)
\label{basic_gamma}
\end{eqnarray}
where
\begin{equation}
x(\mu)\equiv \frac{\alpha_s(\mu)}{\pi}\quad , \quad
w=w(\mu,m_b(\mu))\equiv \frac{m_b^2(\mu)}{\mu^2}\quad ,\quad  
K\equiv \frac{G_F^2\left| V_{ub}\right|^2}{192\pi^3}\quad .
\label{gamma_defns}
\end{equation}
If $m_b$ is identified with a scale-independent pole mass, the known first and second order terms 
within (\ref{basic_gamma}) appear to be poorly convergent because of the proximity of a Borel-plane singularity \cite{vanRit,Borel}.  In ref. \cite{vanRit}, this problem is averted by employing $\overline{{\rm MS}}$ renormalization in which $m_b$ is identified with a 
scale-dependent (running) $b$-quark mass for four or five active flavours.  The one- and two-loop order 
$\overline{{\rm MS}}$ coefficients within 
(\ref{basic_gamma}) are then given by \cite{vanRit}
\begin{eqnarray}
& &{\rm all}~ n_f:\quad a_0=4.25360\quad ,\quad a_1=5\quad , \nonumber\\
& & n_f=5:\quad b_0=26.7848\quad ,\quad b_1=36.9902\quad ,\quad b_2=17.2917\quad ,\nonumber \\
& & n_f=4: \quad b_0=25.7547\quad ,\quad b_1=38.3935\quad ,\quad b_2=17.7083\quad .
\label{pert_coeffs}
\end{eqnarray}

     We utilize asymptotic Pad\'e-approximant methods in the present note to extract an estimate of the three-loop order contributions to (\ref{basic_gamma}).  Such estimates are then tested against explicit ($\overline{{\rm MS}}$) renormalization-group (RG) determinations of 
$c_1$, $c_2$, and $c_3$ (the coefficient $c_0$ is RG-inaccessible to order $x^3$).  We also find, as in \cite{vanRit}, that the 
optimal value of the scale parameter $\mu$ (determined by scale-insensitivity)  for evaluating (\ref{basic_gamma}) is substantially below $m_b(m_b)$ 
[defined to be the solution to $\mu= m_b(\mu)$]. We then present an estimate of the magnitude and theoretical uncertainties of the 
inclusive semileptonic $b\rightarrow u$ decay rate (\ref{basic_gamma}).
    
 To begin, consider a truncated perturbative series
\begin{equation}
S_{N+1}=1+\sum_{n=1}^{N+1} R_n x^n\quad ,\quad 
S\equiv \lim_{N\rightarrow\infty} S_{N}
\label{basic_series}
\end{equation}
in which  $\left\lbrace R_1,~ R_2, \ldots ,~ R_{N+1}\right\rbrace$ are known.  We seek to utilize Pad\'e approximant methods to 
estimate the first unknown term $R_{N+2}x^{N+2}$ within the series $S$.  For example, the set of coefficients  
 $\left\lbrace R_1,~ R_2, \ldots ,~ R_{N+1}\right\rbrace$
is sufficient to generate an $[N \vert 1]$ approximant to the series $S$,
\begin{equation}
S_{[N\vert1]}=\frac{1+a_1x+\ldots +a_N x^N}{1+b_1 x} \quad ,
\label{N_1_pade}
\end{equation}
such that the first $N+2$ terms of the Maclaurin expansion for $S_{[N\vert 1]}$ correspond to the truncated series $S_{N+1}$.  The next term of this Maclaurin expansion represents a Pad\'e-approximant estimate of $R_{N+2}x^{N+2}$. Specifically, we find that
\begin{equation}
R_{N+2}^{[N\vert1]}=\frac{R^2_{N+1}}{R_N} \quad .
\label{PAR}
\end{equation}
To improve such an estimate, we note that the relative error of an $[N| M]$-approximant estimate [as in (\ref{PAR})] to terms in a 
perturbative field-theoretical series is anticipated to be  \cite{Pade1,egks,gardi,begks,Pade2}
\begin{equation}
\delta_{N+M+1}^{[N\vert M]}\equiv \frac{R_{N+M+1}^{[N\vert M]} -R_{N+M+1}}{R_{N+M+1}}=-\frac{M! A^M}{\left(N+M(1+a)+b\right)^M}\quad .
\label{pade_err} 
\end{equation}
In (\ref{pade_err}), $R^{[N\vert M]}_{N+M+1}$   is the estimate for $R_{N+M+1}$ obtained from a Maclaurin expansion of the $[N\vert M]$ approximant generated by the first $N+M$ terms of a series $S$.  Consequently, $\delta_{N+M+1}$ is the relative error 
associated with such a Pad\'e estimate, and the constants $A$, $a$, and $b$ serve to parameterize this relative error. We now use 
(\ref{pade_err}) to obtain an improved estimate of the three-loop coefficient $R_3$ in a series, given knowledge only of the one- 
and two-loop coefficients $R_1$ and $R_2$. If $M = 1$ and if we assume (as in prior applications \cite{Pade2,Pade3}) that $a+b = 0$, 
we see from substitution of (\ref{PAR}) into (\ref{pade_err}) that
\begin{eqnarray}
& &\delta_2^{[0\vert 1]}=\frac{R_1^2-R_2}{R_2}=-A \quad ,
\label{delta_2}\\
& &\delta_3^{[1\vert 1]}=\frac{R_2^2/R_1-R_3}{R_3}=-A/2 \quad .
\label{delta_3}
\end{eqnarray}
Substituting (\ref{delta_2}) into (\ref{delta_3}), we see that \cite{our_pade}
\begin{equation}
R_3=\frac{2R_2^3}{R_1\left(R_1^2+R_2\right)}\quad .
\label{APAP}
\end{equation}
For the perturbative series on the right-hand side of (\ref{basic_gamma}), both the known coefficients $R_{1,2}$  and the unknown coefficient $R_3$ are explicitly of the form [see (\ref{basic_gamma})]:
\begin{eqnarray}
& &R_1(w)=a_0-a_1\log(w) \quad ,
\label{R_1}\\
& & R_2(w)=b_0-b_1\log(w)+b_2\log^2(w) \quad ,
\label{R_2}\\
& & R_3(w)=c_0-c_1\log(w)+c_2\log^2(w)-c_3\log^3(w) \quad .
\label{R_3}
\end{eqnarray} 
To obtain the unknown coefficients $c_{0-3}$ within the three-loop order term $R_3(w)$, we compare moments of $R_3(w)$ over the 
ultraviolet region $0 < w \le 1$ [{\it i.e.}, $\mu\ge  m_b(m_b)$], as obtained from substitution of (\ref{R_1}) and (\ref{R_2}) into (\ref{APAP}), to the same moments obtained explicitly from (\ref{R_3}):
\begin{eqnarray}
N_k&=&(k+2)\int\limits_0^1 dw\, w^{k+1}\left[ \frac{2R_2^3(w)}{R_1(w)\left(R_1^2(w)+R_2(w)\right)}  \right]
\nonumber\\
&=&(k+2)\int\limits_0^1dw\,w^{k+1}\left[  c_0-c_1\log(w)+c_2\log^2(w)-c_3\log^3(w)\right]\quad .
\label{basic_moments}
\end{eqnarray}
Using the $n_f = 5$ values given in (\ref{pert_coeffs}) for the known coefficients $\left\lbrace a_{0-1},b_{0-2}\right\rbrace$, we obtain the 
following set of equations:
\begin{eqnarray}
N_{-1}=1118.64=c_0+c_1+2c_2+6c_3 \quad ,
\label{Nm1}\\
N_0=457.181=c_0+\frac{1}{2} c_1+\frac{1}{2} c_2 +\frac{3}{4}c_3 \quad ,
\label{N0}\\
N_1=337.233 = c_0 +\frac{1}{3}c_1+\frac{2}{9} c_2+\frac{2}{9} c_3 \quad ,
\label{N1}\\
N_2=291.645=c_0+\frac{1}{4} c_1+\frac{1}{8} c_2+\frac{3}{32} c_3\quad .
\label{N2}
\end{eqnarray}
The solution of these equations are our estimates of the three-loop coefficients in (\ref{basic_gamma}) for the case of five 
active flavours:  
\begin{equation}
n_f = 5 (est.): c_0 = 200.5~,\quad c_1 = 251.4~, \quad c_2 = 190.5~, \quad c_3 = 47.61
\label{5fc}
\end{equation}
True values for the coefficients $c_{1-3}$ (but not $c_0$) can be extracted from the invariance of the physical transition rate 
(\ref{basic_gamma}) under changes of scale $\mu$ \cite{vanRit},
\begin{equation}
0=\mu\frac{d}{d\mu}\Gamma\left(\mu,m_b(\mu),x(\mu)\right)=
\left[\mu\frac{\partial}{\partial\mu}-\left(\gamma_0 x+\gamma_1x^2+\gamma_2 x^3+\ldots \right)
m_b\frac{\partial}{\partial m_b}-\left(\beta_0x^2+\beta_1 x^3+\ldots\right)\frac{\partial}{\partial x}\right] \Gamma\quad ,
\label{RG}
\end{equation}
an equation whose order-$x^3$ contributions vanish provided
\begin{eqnarray}
& &c_1=2b_0\beta_0+a_0\beta_1+\gamma_0\left(5b_0-2b_1\right)+\gamma_1\left(5a_0-2a_1\right)+5\gamma_2\quad ,
\nonumber \\
& &c_2=\frac{1}{2}\left[2b_1\beta_0+a_1\left(\beta_1+5\gamma_1\right)+\gamma_0\left( 5 b_1-4 b_2\right)\right] \quad ,
\label{RG_c}\\
& &c_3=\frac{b_2}{3}\left(2\beta_0+5\gamma_0\right)\quad .
\nonumber
\end{eqnarray}
Using the $n_f = 5$ values in (\ref{pert_coeffs}) and the values $\beta_0=23/12$, $\beta_1 =29/12$,  $\gamma_0 = 1$,  
$\gamma_1 = 253/72$, and  $\gamma_2 = 7.41985$ for five active flavours, we find from (\ref{RG_c}) that
\begin{equation}
n_f = 5: \quad c_1 = 249.592~,\quad c_2 = 178.755~, \quad c_3 = 50.9144
\label{5fRGc}
\end{equation}
The estimates (\ref{5fc}) for $c_{1-3}$ are all within 7\% of these correct values.\footnote{ 
The moment methodology delineated above has 
also been employed with similar success to the RG-accessible order-$x^3$ terms of the gluonic scalar-current correlation function  
\protect\cite{Pade_RG1}
and the CP-odd Higgs boson decay rate \protect\cite{Pade_RG2}.}

     Since we do not know the true value of $c_0$, it  is important  to test whether the value obtained in (\ref{5fc}) 
is stable 
when the estimates of $c_{1-3}$ derived from (\ref{Nm1}--\ref{N2}) are replaced with the true values given in (\ref{5fRGc}). If we utilize (\ref{5fRGc}) explicitly within (\ref{Nm1}--\ref{N2}), we obtain four separate determinations of $c_0$, which are 
respectively given by $c_0 = 206.0$, 204.8, 203.0 and 202.1. These results are all within 1\% of each other, and 1--3\% above 
the estimate obtained in (\ref{5fc}).  A further consistency check on the $c_0$ estimate is to construct a least-squares fit of 
equation (\ref{APAP}) to the form of eq. (\ref{R_3}) over the entire region $\mu \ge m_b(m_b)$: 
\begin{equation}
\chi^2\left(c_0,c_1,c_2,c_3\right)=\int\limits_0^1dw\left[
\frac{2R_2^3(w)}{R_1(w)\left(R_1^2(w)+R_2(w)\right)}-
\left( c_0-c_1\log(w)+c_2\log^2(w)-c_3\log^3(w)  \right)
\right]^2 \quad .
\label{chi2}
\end{equation}
The minimization requirements  $\partial\chi^2/\partial c_i = 0$ lead to the following estimates for $c_i$:
\begin{equation}
n_f = 5 (est.):\quad c_0 = 198.4~,\quad c_1 = 260.6~,\quad c_2 = 183.9~,\quad c_3 = 48.64~,\quad  \chi^2_{min} = 1.505
\label{5fchi2c}
\end{equation}
Note that $c_{1-3}$ are all within 4.5\% of their true values (\ref{5fRGc}), and that the small value for  $\chi^2_{min}$ implies a 
precise fit of (\ref{APAP}) to the form of (\ref{R_3}) over the entire interval $0 < w \le 1$.  The estimates of $c_0$ obtained in 
(\ref{5fchi2c}) and (\ref{5fc}) are remarkably close. If true values (\ref{5fRGc}) for $c_{1-3}$ are substituted directly into  $\chi^2$,
minimization with respect to the only remaining variable $c_0$ leads to the estimate
\begin{equation}
n_f = 5 (est.):\quad c_0 = 206.0\quad , 
\label{5f_bestc0}
\end{equation}
identical to that obtained by a similar substitution into (\ref{Nm1}). We shall focus on this particular estimate when estimating the 
rate (\ref{basic_gamma}), despite the fact that it is 3--4\% above the estimates in (\ref{5fc}) and (\ref{5fchi2c}), because this estimate incorporates the RG-determinations of values for $c_{1-3}$.
  
   The procedures delineated above can also be applied to the case of four active flavours.  Utilization of the $n_f = 4$ values for 
$\left\lbrace a_{0-1},~b_{0-2}\right\rbrace$, as tabulated in (\ref{pert_coeffs}), in the first integrand of (\ref{basic_moments}) leads to the 
following numerical values for the moments $N_k$: 
\begin{equation}
N_{-1} = 1165.14~,\quad N_0 = 457.864~,\quad N_1 = 329.529~,\quad N_2 = 280.696\quad . 
\label{4fmoments}
\end{equation}
If these values are incorporated into (\ref{Nm1}--\ref{N2}), the solution to the resulting set of equations is  
\begin{equation}
n_f = 4 (est.):\quad c_0 = 182.7~,\quad c_1 = 271.9~,\quad c_2 = 201.4~,\quad c_3 = 51.26
\label{4fc}
\end{equation} 
True values for $c_{1-3}$ are RG-accessible via eqs. (\ref{RG_c}), with $n_f = 4$ values for $\left\lbrace a_{0-1},~b_{0-2}\right\rbrace$ as listed in (\ref{pert_coeffs}), and with  $\beta_0= 25/12$, $\beta_1 =77/24$, $\gamma_0 = 1$, $\gamma_1 = 263/72$, and  $\gamma_2 = 9.94702$ for four active flavours. The correct values are
\begin{equation}
n_f = 4:\quad c_1 = 263.84~,\quad c_2 = 194.23~,\quad c_3 = 54.109
\label{4fRG}
\end{equation}
which are within 5.3\% of their estimated values in (\ref{4fc}). As before, the incorporation of the correct values (\ref{4fRG}) into the moment equations (\ref{Nm1}--\ref{N2}) with numerical values (\ref{4fmoments}) for $n_f = 4$ moments leads to four separate estimates for $c_0$ $\left\lbrace188.2,~ 188.3,~ 186.4,~ 185.4\right\rbrace$ that are remarkably consistent, and all less than 
3\% above the estimate in (\ref{4fc}).  Moreover, minimization of the  $\chi^2$ function (\ref{chi2}) using (\ref{R_1}) and (\ref{R_2}) 
with $n_f = 4$ values for $\left\lbrace a_{0-1},~b_{0-2}\right\rbrace$ yields estimates for $c_{1-3}$ all within 5.1\% of their true values 
(\ref{4fRG}); {\it i.e.,}   $\partial \chi^2/\partial c_i = 0$ when
\begin{equation}
n_f = 4 (est.):\quad c_0 = 181.5~,\quad c_1 = 277.3~,\quad c_2 = 197.6~,\quad c_3 = 51.86~,  \quad \chi^2_{min} = 0.5079
\label{4fchi2c}
\end{equation}
The above estimate for $c_0$ is within 1\% of that obtained in (\ref{4fc}). If one incorporates into  $\chi^2$ the true values 
(\ref{4fRG}) for $c_{1-3}$, one finds that   $\partial\chi^2/\partial c_0= 0$ when
\begin{equation}
n_f = 4 (est.):\quad c_0 = 188.2\quad,
\label{4f_bestc0}
\end{equation}
analogous to the estimate (\ref{5f_bestc0}) for five active flavours. 

To examine the consistency of the Pad\'e-approximant 
methodology leading to (\ref{4f_bestc0}), one can obtain an estimate of the $n_f = 4$ value for $c_0$ directly from (\ref{5f_bestc0}) via comparison of four- and five-active-flavour expressions for the decay rate (\ref{basic_gamma}).  Suppose for $n_f = 5$ 
[indicated henceforth by the superscript $\phantom{~}^{(5)}$] that we evaluate (\ref{basic_gamma}) at $\mu = 4.17\,{\rm GeV}$ in 
order to coincide with (the central value of) an empirical estimate of $m_b^{(5)}(m_b)$ \cite{mb_ref}, and we choose 
$x^{(5)}(4.17 {\rm GeV}) = 0.0715492$, a value devolving (numerically) via the four-loop $\beta$-function from 
$\alpha_s^{(5)}(M_Z) = 0.118$ 
(the corresponding  input value in \cite{mb_ref}). If we substitute into (\ref{basic_gamma}) the $n_f = 5$ estimate 
(\ref{5f_bestc0}) for $c_0$ and known $n_f = 5$ values (\ref{pert_coeffs}) for $a_0$ and $b_0$, and if we note that 
$\log(w) = 0$ when $\mu = m_b^{(5)}(\mu)$, the defining relationship for $m_b^{(5)}(m_b)$, we find that
\begin{equation}
\frac{\Gamma^{(5)}}{K} = (4.17\, {\rm GeV})^5\left[1 + 0.304342 + 0.137119 + 0.075454\right] = 1912.68\, {\rm GeV}^5   
\label{5fGamma}
\end{equation}
The same rate can be evaluated for the case of four active flavours with $c_0$ left arbitrary by substituting into (\ref{basic_gamma}) the $n_f = 4$ RG values (\ref{4fRG}) for $c_{1-3}$ and the $n_f = 4$ values (\ref{pert_coeffs}) for $\left\lbrace a_{0-1},~b_{0-2}\right\rbrace$, and by utilizing following threshold-matching conditions \cite{match} to obtain values for $x(4.17\, {\rm GeV})$ and $m_b(4.17\,{\rm GeV})$ appropriate for four active flavours, as indicated by the superscript $\phantom{~}^{(4)}$:
\begin{eqnarray}
x^{(4)}(4.17\,{\rm GeV})&=&x^{(5)}(4.17\,{\rm GeV})\left[1+0.1528\left[x^{(5)}(4.17\,{\rm GeV})\right]^2
+0.633\left[x^{(5)}(4.17\,{\rm GeV})\right]^3\right]
\nonumber\\
&=&0.0716218\quad ,
\label{match_alpha}\\
m_b^{(4)}(4.17\,{\rm GeV})&=&m_b^{(5)}(4.17\,{\rm GeV})\left[1+0.2060\left[x^{(5)}(4.17\,{\rm GeV})\right]^2+
1.9464\left[x^{(5)}(4.17\,{\rm GeV})\right]^3\right]
\nonumber\\
&=&4.17739\,{\rm GeV}
\label{match_mb}
\end{eqnarray}
We note that $\log(w) =  2 \log\left[m_b^{(4)}(\mu)/\mu\right]$ is no longer zero, and we obtain from (\ref{basic_gamma}) the following 
four-active-flavours expression for the rate:
\begin{equation}
\frac{\Gamma^{(4)}}{K}=\left(1824.8+0.46737 c_0\right)\, {\rm GeV^5}
\label{4fGamma}
\end{equation}
Equating (\ref{5fGamma}) to (\ref{4fGamma}), we find that $c_0$ in the latter expression is 188.03, a value in startlingly close 
agreement with the independent estimate (\ref{4f_bestc0}).  Thus, at the five-flavour threshold $m_b^{(5)}(m_b) = 4.17 \, {\rm GeV}$, 
equilibration of the below-threshold and above-threshold expressions for the inclusive semileptonic $b\rightarrow   u$ rate is seen to 
corroborate the consistency of the independent $c_0$ estimates (\ref{4f_bestc0}) and (\ref{5f_bestc0}) for four and 
five active flavours. 

     As noted in \cite{vanRit}, the series (\ref{basic_gamma}) will exhibit some $\mu$ dependence as an artifact of its truncation; 
{\it i.e.}, the three loop expression for $\Gamma$ is subject to ${\cal O}(x^4)$ violations of (\ref{RG}).  In 
Figure 1, we display this residual $\mu$ dependence of the rate $\Gamma[\mu]/K$ both below and above the 
five-flavour threshold, which is assumed (as above) to occur at $m_b^{(5)}(m_b) = 4.17 \, {\rm GeV}$. For $\mu$ greater than this 
threshold, we obtain $\Gamma[\mu]/K$ utilizing known $n_f = 5$ values (\ref{pert_coeffs},\ref{5fRGc}) for 
$\left\lbrace a_{0-1},b_{0-2},c_{1-3}\right\rbrace$ in conjunction with the estimate (\ref{5f_bestc0}) for $c_0$. The above-threshold values for 
$m_b^{(5)}(\mu)$ and $x^{(5)}(\mu)$ are evolved numerically via $n_f = 5$ four-loop RG-functions from the specific values $m_b^{(5)}(4.17 \, {\rm GeV}) = 4.17 \, {\rm GeV}$ and $x^{(5)}(M_Z) = 0.118/\pi$. For $\mu$ less than $4.17 \, {\rm GeV}$, we utilize 
known $n_f =4$ values (\ref{pert_coeffs},\ref{4fRG}) for $\left\lbrace a_{0-1},b_{0-2},c_{1-3}\right\rbrace$ in conjunction with the estimate 
(\ref{4f_bestc0}) for $c_0$. The below-threshold values for $x^{(4)}(\mu)$ and $m_b^{(4)}(\mu)$ are evolved via $n_f = 4$ four-loop 
RG-functions from the specific values for $x^{(4)}(4.17 \, {\rm GeV})$ and $m_b^{(4)}(4.17 \, {\rm GeV})$ obtained in 
(\ref{match_alpha}) and (\ref{match_mb}). We note that the maximum of the rate exhibited in Fig. 1 occurs at 
 $\mu = 1.775 \, {\rm GeV}$ (near the $\tau$ mass), and that
\begin{equation}
\frac{\Gamma[1.775 \, {\rm GeV}]}{K} = \left[ 5.1213 \, {\rm GeV}\right]^5 \left[1 - 0.6455 + 0.2477 - 0.0143\right] 
          = 2071 \, {\rm GeV}^5\quad . 
\label{Gamma_opt}
\end{equation}
This value corresponds to the minimal-sensitivity \cite{stevenson} prediction of the decay rate.

     In order to get a handle on the theoretical uncertainty of this estimate, we will assume that there exists a bi-directional series truncation error equal to the three-loop contribution to (\ref{Gamma_opt}). An additional source of theoretical uncertainty is 
the error associated with the estimate (\ref{4f_bestc0}) for $c_0$. This uncertainty should be comparable to the explicit error in estimating $c_1$, $c_2$, and $c_3$ by the same least squares fitting procedure in (\ref{4fchi2c}). Comparing the true values (\ref{4fRG}) to these estimates, we have already noted that such errors are 5\% or less, a value we shall assume to 
characterize error in the 
estimate (\ref{4f_bestc0}). Such an error estimate is upheld by the range  $198<c_0<206$  obtained in the different methodological approaches to determining $c_0$ [see (\ref{5fc},\ref{5f_bestc0})].
As noted above, the estimate (\ref{Gamma_opt}) is based upon (correlated \cite{mb_ref}) input values 
$\alpha_s(M_Z) = 0.118$ and $m_b^{(5)}(m_b) = 4.17 \, {\rm GeV}$.  The errors associated with these estimates are 
$\pm 0.004$ and $\pm 0.05 \, {\rm GeV}$, respectively \cite{mb_ref}. Finally, we quote the ref. \cite{non_pert} estimate of the 
nonperturbative (NP) contribution to $\Gamma/K$:
\begin{equation}
\left(\frac{\Delta\Gamma}{K}\right)_{NP}=m_b^3\left(\frac{-9\lambda_2+\lambda_1}{2}\right)\quad ,\quad
-0.5\,{\rm GeV^2}\le\lambda_1\le 0\quad ,\quad \lambda_2=0.12\,{\rm GeV^2}
\label{non_pert}
\end{equation}
Taking into account all of these uncertainties, we estimate that
\begin{equation}
\frac{192\pi^3\Gamma(b\rightarrow u\bar\nu_\ell \ell^-)}{G_F^2\left\vert V_{ub}\right\vert^2}=
2071\,{\rm GeV^5}\pm 51\,{\rm GeV^5} \pm 35 \,{\rm GeV^5} \left(^{+144}_{-112}\right) \,{\rm GeV^5}
\left(^{+120}_{-115}\right) \,{\rm GeV^5}
\left(^{-39}_{-57}\right) \,{\rm GeV^5}
\label{final_gamma}
\end{equation} 
where the listed set of uncertainties respectively correspond to uncertainties associated with series truncation, 
the $c_0$ estimate, the value for $\alpha_s\left(M_Z\right)$, the value for
$m_b^{(5)}(m_b)$, and the nonperturbative contribution (\ref{non_pert}).
 
     The truncation error listed in (\ref{final_gamma}) may well be excessive. The series in (\ref{Gamma_opt}) appears to be an 
{\em alternating} series whose final term is quite small.  If this alternating character is retained for higher orders of $x$, then 
the series within (\ref{Gamma_opt}) is bounded from above by the sum of its first three terms and bounded from below by the 
sum of all four known terms, in which case the truncation error should be unidirectional:
\begin{equation}
               2071 \, {\rm GeV}^5 \le \left(\frac{\Gamma}{K}\right)_{true}\le  2122 \, {\rm GeV}^5
\label{bounds}
\end{equation}
The above suggests  that truncation error at this point of minimum scale-sensitivity may lead to a spread as small
as $\pm 1.3\%$.
However, the smallness of the order-$x^3$ contribution suggests that the series (\ref{Gamma_opt}) may be asymptotic, consistent with the
anticipated behaviour ($R_N\sim N! B^{-N}N^c$ \cite{asymptotic}) underlying eq. (\ref{pade_err}).  If such is the case,
the magnitude of the order-$x^4$-and-higher terms may well increase with order.  Since the value for a quantity represented by 
an asymptotic series is obtained by a sum of only its decreasing terms (with an error comparable to the smallest such term), an optimal 
determination of $\left(\Gamma/K\right)_{true}$ may well be obtained by choosing $\mu$ such that the order-$x^3$ term {\em vanishes}.
This value of $\mu=1.835\,{\rm GeV}$ corresponds to the point of Figure 2 at which the two-loop and three-loop expressions for $\Gamma/K$
cross:
\begin{equation}
\frac{\Gamma(1.835\,{\rm GeV})}{K}=2069\,{\rm GeV^5}
\label{equiv_Gamma}
\end{equation}

     The estimates (\ref{Gamma_opt}) and (\ref{equiv_Gamma}) are 8\% larger than the rate (\ref{5fGamma}) at the five-flavour threshold.  The former values represent a superior estimate for the true rate, since the rate at this point is locally 
insensitive to the renormalization scale parameter $\mu$. The discrepancy between (\ref{5fGamma}) and (\ref{Gamma_opt}) can be 
understood by noting that (\ref{5fGamma}) appears to be a positive term series.  If convergent, a truncated positive term series is 
necessarily an underestimate of the series sum. Indeed, the rate displayed in Fig. 1 falls off even more 
(as noted in \cite{vanRit}) for values of $\mu$ above the five-flavour threshold $m_b^{(5)}(m_b)$.  This progressive fall off 
can be understood by noting that the entire set of coefficients $\left\lbrace a_{0-1},b_{0-2},c_{0-3}\right\rbrace$ is positive, and that when 
$\mu$ increases above the five-flavour threshold, $\log(w)$ becomes increasingly negative. Consequently, we see from (\ref{basic_gamma}) that the 
coefficients of $x$, $x^2$, and $x^3$ become increasingly positive, more than offsetting the decrease in $x$ itself with increasing
 $\mu$.  This serves to increase the magnitude of the successive terms within the positive term series, which necessarily implies an 
increase in the truncation error, as well  as a progressive {\em underestimation} of the full rate, as a consequence of truncation to three-loop order. 

     One way to obviate these difficulties and to recover (approximate) scale invariance of the full series for the rate $\Gamma$ is 
to estimate this sum via the use of Pad\'e approximants. For example, such Pad\'e-summation of 
the known series within the Bjorken sum rule has been shown \cite{egks} to reduce substantially the scale dependence of the sum-rule-extracted value for $\alpha_s$.  This reduction in scale dependence via the substitution of (diagonal or near-diagonal) Pad\'e approximants in place of truncated field-theoretical perturbative series has been more generally established in reference \cite{gardi}.

For the case of the inclusive semileptonic $b\to u$ decay rate, we display in Figure 3 the three-loop truncated rate together with its corresponding $[1|2]$ and $[2| 1]$ approximants [{\it i.e.}, the $[1| 2]$ and $[2| 1]$ Pad\'e-summations of the truncated 
perturbative series $S_3$, as defined by (\ref{basic_series})]. The approximants are seen to exhibit substantially less $\mu$-dependence than the three loop expression for the rate, and are seen to be reasonably close to the minimal-sensitivity value 
obtained in (\ref{Gamma_opt}). The $[2| 1]$ approximant to the rate remains between $2001\, {\rm GeV}^5$ and 
$2029 \, {\rm GeV}^5$ over the range 
$4.17 \, {\rm GeV} \le \mu\le  9 \, {\rm GeV}$, and the $[1| 2]$ approximant remains between $1956\, {\rm GeV}^5$ and 
$2053 \, {\rm GeV}^5$ over the 
same range.  By contrast, the truncated series itself is seen from Figure 3 to fall from $1913 \, {\rm GeV}^5$ at 
$\mu = 4.17 \, {\rm GeV}$ to  $1733 \, {\rm GeV}^5$ when $\mu = 9\,\,{\rm GeV}$.

At present, the experimental uncertainties in direct measurements of the
inclusive semileptonic $b \rightarrow u$ decay rate are substantially
larger than the theoretical uncertainties listed in (37); {\it e.g.},
recent ALEPH \cite{aleph} and L3 \cite{l3} Collaboration branching-ratio 
measurements are $(1.73 \pm 0.78)\cdot 10^{-3}$ and $(3.3
\pm2.0)\cdot10^{-3}$, respectively, with systematic and statistical
errors combined via quadrature.  Somewhat less direct measurements based
upon endpoint region measurements of the $\Upsilon$(4S) decay spectrum are
characterized by additional model dependence, although they are utilised
as a source of present experimental bounds \cite{caso} on
$V_{ub}:|V_{ub} / V_{cb}|=0.08 \pm 0.02$ and $|V_{ub}|=0.0037 \pm
0.0006$.  However, as experimental uncertainties in the semileptonic $b
\rightarrow u$ rate are reduced, control over the uncertainty
within theoretical expressions for the rate becomes increasingly
important.\footnote{It has recently been demonstrated \cite{mel} that four
differing literature-motivated estimates of the three-loop coefficient
$c_0$ (including our own) can themselves result in a 7\% spread in the
theoretical rate, without taking into consideration any of the other
sources of uncertainty listed in (37).} The estimate we present above for
the unknown three-loop coefficient $c_0$ is corroborated both by concomitant
success in predicting the RG-accessible three-loop terms within the
theoretical rate (1), as well as by consistency between the direct 
four-active-flavour (asymptotic-Pad\'e) determination of $c_0$ (30) and 
its independent determination via (25) and (34) [i.e. via
flavour-threshold matching conditions].  Given the total b-decay rate
$\Gamma_{total}=(4.249\pm0.039)\cdot10^{-13}$ GeV \cite{L3C}, our result
(37) leads to the following relationship between $|V_{ub}|$ and the
inclusive semileptonic branching ratio (BR):

\begin{eqnarray*}BR=|V_{ub}|^2 \left(\frac{G_F^2}{192\pi^3}\right)
\frac{[(2071_{pert.}-48_{nonpert.})GeV^5 \pm 17\%]}{4.25\cdot10^{-13}GeV},
\end{eqnarray*}
\begin{equation}
|V_{ub}|=0.0959\cdot(BR)^{1/2}\pm9\%.
\end{equation}

\noindent
{\bf Acknowledgments:} VE and TGS gratefully acknowledge research funding from the Natural Sciences and Engineering Research Council of Canada (NSERC). VE wishes to thank P.J. Sullivan for a useful discussion.

\begin{figure}[hbt]
\centering
\includegraphics[scale=0.5,angle=270]{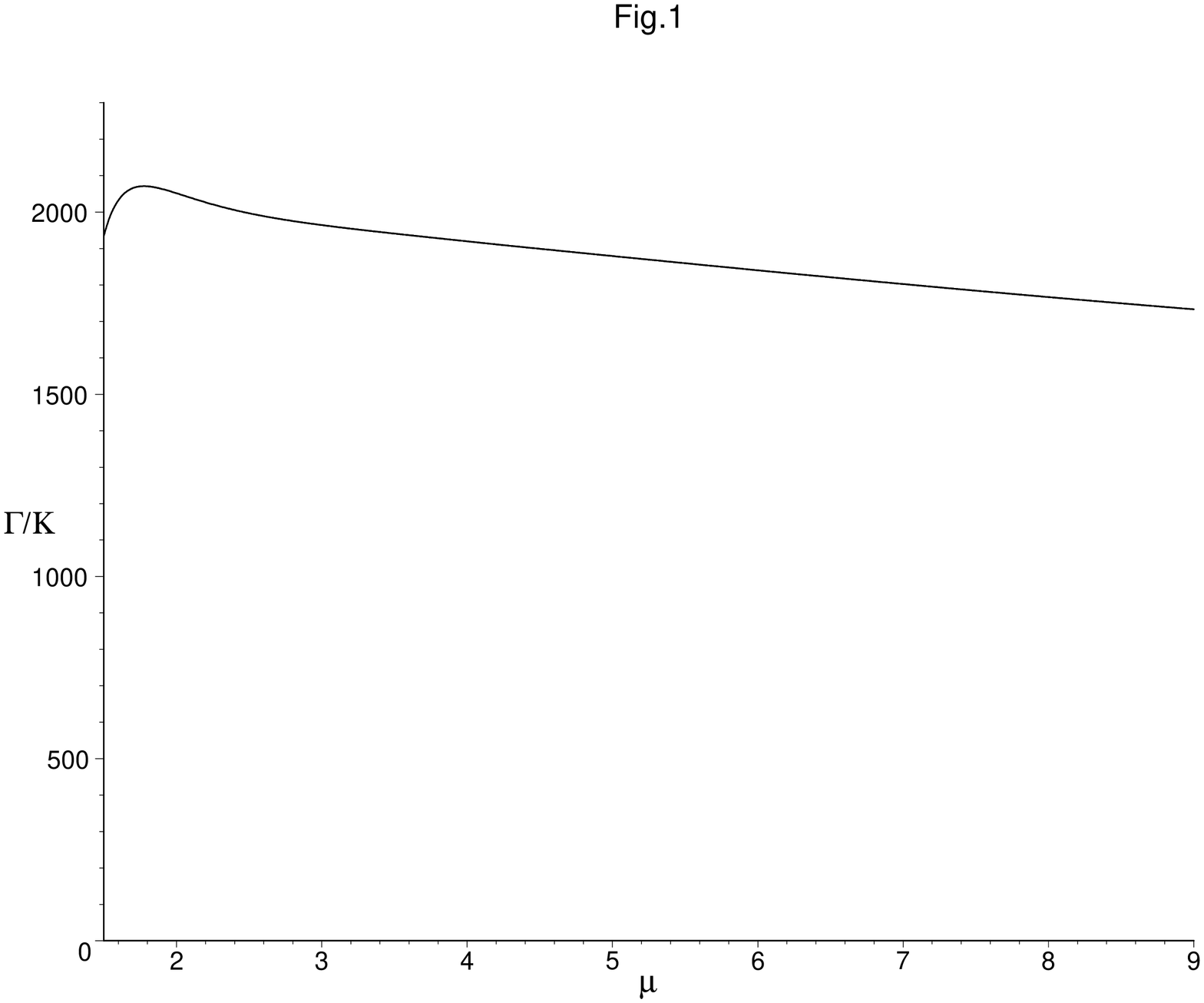}
\caption{
The $\mu$-dependence of the three-loop order decay rate $\Gamma(b\rightarrow u \bar\nu_\ell\ell^-)/K$ obtained from
input values $\alpha_s(M_Z) = 0.118$, $m_b(m_b) = 4.17\, {\rm GeV}$. The rate is calculated using four active flavours
below and five active flavours above the five-flavour threshold, subject to threshold matching
conditions discussed in the text.  All scales are in GeV units.
}
\label{Gamma_vs_mu}
\end{figure}

\begin{figure}[hbt]
\centering
\includegraphics[scale=0.5,angle=270]{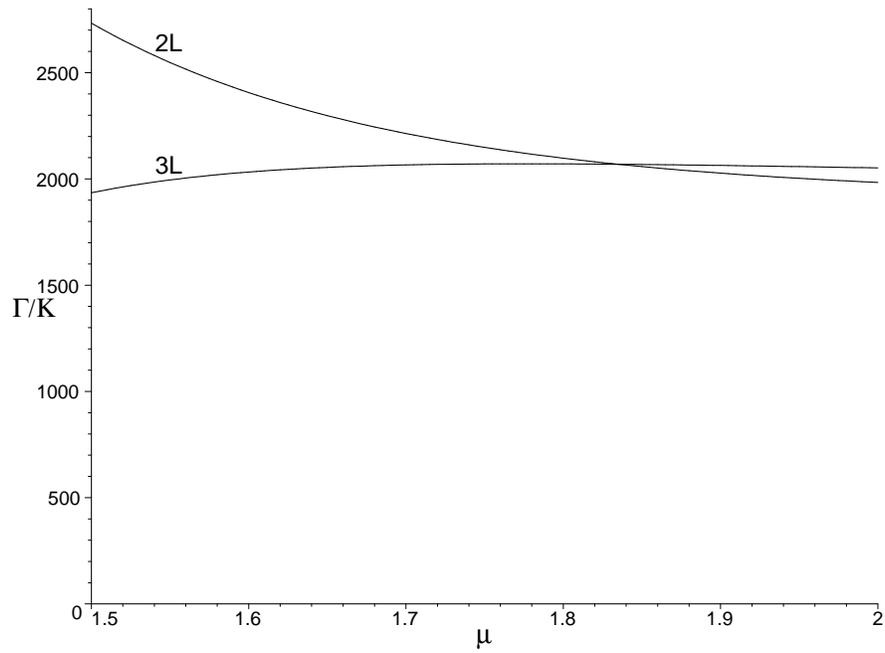}
\caption{
The two-loop (2L) and three-loop (3L) expressions for  $\Gamma(b\rightarrow u \bar\nu_\ell\ell^-)/K$ obtained from
input values $\alpha_s(M_Z) = 0.118$, $m_b(m_b) = 4.17 \,{\rm GeV}$ in the region of minimal $\mu$-sensitivity of the
three-loop rate.  Both curves are obtained using four active flavours.  All scales are in GeV units.
}
\label{2_3_loop}
\end{figure}

\begin{figure}[hbt]
\centering
\includegraphics[scale=0.5,angle=270]{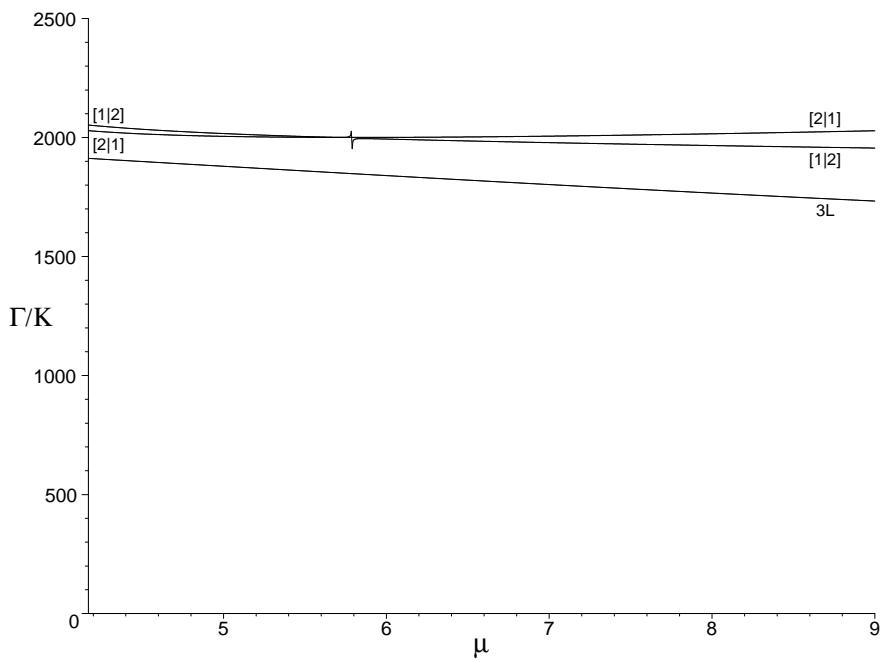}
\caption{
 Comparison of three-loop (3L) and Pad\'e-summation expressions for the rate  $\Gamma(b\rightarrow u \bar\nu_\ell\ell^-)/K$
obtained from input values $\alpha_s(M_Z) = 0.118$, $m_b(m_b) = 4.17 \,{\rm GeV}$ when $\mu$ is above
the five-flavour threshold. The summations are $[1|2]$ and $[2|1]$ approximants whose Maclaurin
expansions reproduce the 3L expression.  All scales are in GeV units.
}
\label{pade_sum}
\end{figure}

\end{document}